\documentclass[preprint,11pt,preprintnumbers,nofootinbib]{revtex4}
\usepackage{graphicx}
\usepackage{dcolumn}
\usepackage{bm}
\usepackage{amsfonts}

\begin{document}

\date{\today}
\title{Staggered Multi-Field Inflation}

\author{Diana Battefeld} 
\email[email: ]{diana.battefeld(AT)helsinki.fi}
\affiliation{Helsinki Institute of Physics, P.O. Box 64, University of Helsinki, FIN-00014 Helsinki, Finland}
\author{Thorsten Battefeld}
\email{tbattefe(AT)princeton.edu}
\affiliation{ Princeton University,
Department of Physics,
NJ 08544
}
\author{Anne-Christine Davis}
\email{A.C.Davis(AT)damtp.cam.ac.uk}
\affiliation{DAMTP, Center for Mathematical Sciences, University of Cambridge, Wilberforce Road, Cambridge,CB3 0WA, UK}

\begin{abstract}
We investigate multi-field inflationary scenarios with fields that drop out of the model in a staggered fashion. This feature is  natural in certain multi-field inflationary setups within string theory; for instance, it can manifest itself when fields are related to tachyons that condense, or inter-brane distances that become meaningless when branes annihilate. Considering  a separable potential, and promoting the number of fields to a smooth  time-dependent function, we derive the formalism to deal with these models at the background and perturbed level, providing general expressions for the scalar spectral index and the running.   We recover known results of e.g. a dynamically relaxing cosmological constant in the appropriate limits. We further show that isocurvature perturbations are suppressed during inflation, so that perturbations are adiabatic and nearly Gaussian. The resulting setup might be interpreted as a novel type of warm inflation, readily implemented within string theory and without many of the shortcomings associated with warm inflation.

To exemplify the applicability of the formalism we consider three concrete models: assisted inflation with exponential potentials as a simple toy model (a graceful exit becomes possible), inflation from multiple tachyons (a constant decay rate of the number of fields and negligible slow roll contributions turns out to be in good agreement with observations) and inflation from multiple M5-branes within M-theory (a narrow stacking of branes yields a consistent scenario).
\end{abstract}
\maketitle
\newpage

\tableofcontents

\section{Introduction}
An inflationary epoch of the early universe is widely accepted as the most efficient mechanism to solve the flatness and the horizon problem, while providing a nearly scale invariant spectrum of scalar perturbations, in agreement with observations. However, the embedding of inflation driven by a single scalar field within string theory (presently, the only known self-consistent theory of quantum gravity) has proven to be challenging, since it requires an extremely flat potential. In addition, the presence of many dynamic fields in string theory renders single field models less appealing because all but one degree of freedom need to be already stabilized at the onset of inflation. Though far from being simple or generic, a partially successful implementation is the KKLMMT construction \cite{Kachru:2003sx}.

To ameliorate the problem of fine tuning the potential, Liddle et.~al.~proposed assisted inflation in \cite{Liddle:1998jc}  (see also \cite{Malik:1998gy,Kanti:1999vt} and follow ups), a certain type of multi-field inflation whereby the presence of many fields increases the Hubble friction, allowing for steeper potentials to still drive slow roll inflation. One shortcoming, however, is the need for fine tuned initial conditions for particular potentials, especially in the absence of an attractor solution (for instance in N-flation \cite{Dimopoulos:2005ac}; see \cite{Calcagni:2007sb} for a more general stability analysis of assisted inflation). The presence of many degrees of freedom in string theory has, over the years, galvanized the emergence of a variety of models implementing assisted inflation; these include inflation from multiple tachyons \cite{Majumdar:2003kd}, from multiple M5-branes within M-theory \cite{Becker:2005sg} or from axions \cite{Dimopoulos:2005ac} among others, see e.g. \cite{Cline:2005ty,Ward:2007gs,Grimm:2007hs,Panigrahi:2007sq,Piao:2002vf,Mazumdar:2001mm} for a small selection.

To make predictions in these models, it is common to introduce a single effective degree of freedom, visualized as the length of the trajectory in field space, see e.g. the review \cite{Wands:2007bd}. Utilizing this approach correctly recovers adiabatic perturbations, but not isocurvature/entropy modes. The latter ones can be pictured as perturbations perpendicular to the trajectory and develop whenever several degrees of freedom are present \cite{Gordon:2000hv}. Although non-Gaussianities may be produced (see the review \cite{Wands:2007bd} for details), it is possible to show that they are generically slow roll suppressed \cite{Battefeld:2006sz,Battefeld:2007en} (see however \cite{Byrnes:2008wi}). With this in mind, is it then feasible to discriminate a given multi-field model from its single field analog?

In this article, we investigate a, so far largely ignored (see however \cite{Ashoorioon:2006wc}), discriminating property of multi-field models, namely the possibility that fields decay during inflation. By "decay" we mean that individual fields suddenly become obsolete, while their energy is converted into other forms, such as radiation. This is a widespread feature of models within string theory: consider, for instance, an association of inflatons with distances between branes that are located in some internal space. If a  brane annihilates during inflation, for example as a result of its dissolution into a boundary brane or a collision with an anti-brane, the distance to the just mentioned brane becomes meaningless. Of course, energy does not vanish, but is converted into a different type during the annihilation. A further example is inflation driven by tachyons, which can condense during inflation in a staggered fashion.

The disappearance of fields causes an additional decrease of the potential energy during inflation, which can be even more important than the reduction of  potential energy induced by slow roll. Hence, leading order corrections to observable parameters that are sensitive to the slow roll parameters, such as the scalar spectral index, are possible. Indeed, one can construct models of inflation entirely without a slow roll phase, similar to inflation without inflatons as proposed in \cite{Watson:2006px}.

In order to compute these effects we assume a simple separable potential $W=\sum_{A=1}^{\mathcal{N}}V_A$ and promote the number of fields $\mathcal{N}$ to a time dependent function. We further smooth out $\mathcal{N}$ so that we can introduce a continuous decay rate $\Gamma(t)\equiv -\dot{\mathcal{N}}/\mathcal{N}$. This approach is only justified if the number of fields and $\Gamma$ are large enough so that several fields disappear in any given Hubble time. Further, any signal due to a rapid drop in the potential energy, such as a ringing in the power spectrum  or additional non-Gaussianities \cite{Covi:2006ci,Ashoorioon:2006wc,Chen:2006xjb}, cannot be recovered by this approach.
To guarantee energy-momentum conservation ($\nabla_\nu T^{\mu \nu}=0$), we are forced to introduce an additional component of the total energy momentum tensor that takes over
the energy of the disappearing fields. The ratio of its energy density to the one in the remaining inflatons yields an additional small parameter $\bar{\varepsilon}$ appearing alongside the usual slow roll parameters. As a consequence, the equation of motion of the effective field is modified. The resulting set-up is reminiscent of warm inflation \cite{Berera:1995ie}, and might indeed be seen as a new, less problematic implementation of warm inflation within string theory.

At the perturbed level, we consider adiabatic and entropy perturbations and show that the latter ones are suppressed in the models of interest. Focusing on the Mukhanov variable, we derive the scalar spectral index and its running for general decay rates and number of fields. We recover the usual slow roll result as well as \cite{Watson:2006px} (a dynamically relaxing cosmological constant) in the appropriate limits, but we also find leading order corrections to the slow roll result in general. The running remains second order in small parameters and is, therefore, too small to be observed.

To exemplify the applicability of the formalism we make a detailed study of three concrete models; firstly, as an instructive toy model, we consider assisted inflation with exponential potentials and a nonzero, constant decay rate, which we insert by hand. $\Gamma\neq 0$ provides a graceful exit to inflation at the cost of requiring somewhat flatter potentials or more fields so that the model remains consistent with observations. If the decay rate is too big ($\Gamma\gtrsim 0.04 H_{inf}$) the scalar spectral index becomes unacceptably red, but all smaller rates work well. Secondly, we investigate inflation from multiple tachyons, as proposed in \cite{Majumdar:2003kd}. 
Here, tachyons get displaced from the top of their potential by thermal or quantum mechanical fluctuations, causing them to condense during inflation. We find that the best motivated and least fine tuned setup consists of a constant decay rate and tachyons close to the top of their potential so that their slow roll evolution yields negligible contributions to observable parameters. The resulting spectral index is $n_s-1=3/N$, where $N\approx 60$ is the number efolds at which we evaluate $n_s$. This is in good agreement with the observations of the CMBR, $n_s=0.960^{+0.014}_{-0.013}$ \cite{Komatsu:2008hk}. Lastly, we look at inflation from multiple M-5 branes within M-theroy \cite{Becker:2005sg}. Here, inflatons are associated with the inter-brane distance between M5-branes located within an orbifold $\bf S_1/Z_2$. Whenever a brane comes close to a boundary brane it dissolves into the boundary via a small instanton transition. Given the model's parameter ranges, which encompass, for instance, the concrete potential and the maximal number of branes, we show that inflation comes to an end within a few efolds after the first outermost brane disintegrates. Hence, only an initial narrow stacking of branes  (involves fine tuning) permits a consistent scenario, and the usual slow roll expressions apply. Based upon these findings, cascade inflation as investigated in \cite{Ashoorioon:2006wc} is irrelevant for the cosmological scales that are observed in the CMBR, given the parameters put forward in \cite{Becker:2005sg}.

It is evident that whether or not corrections due to $\Gamma\neq 0$ are important is  model-dependent, ranging from being the primary ingredient, as in the tachyon case, to being not pertinent, as in the M5-brane case. Thus, the many implementations of multi-field inflation within string theory should be thoroughly re-investigated.

The outline of this paper is as follows: In section \ref{sec:main}, we derive the formalism to deal with dacaying fields, at the background level (sec.~\ref{sec:background}) and the perturbed one (sec.~\ref{sec:spectralindex}). We compute the spectral index and the running in section \ref{sec:spectralindex} and \ref{sec:running}. Along the way, we draw comparisons to warm inflation in section \ref{sec:warm} and comment on isocurvature modes as well as non-Gaussianities in section \ref{sec:entropy}. We then shift gears and focus on applications,  that is we investigate concrete models: assisted inflation (sec.~\ref{sec:staggered}), inflation from tachyons (sec.~\ref{sec:ACDMM}) and inflation from multiple M5-branes (sec.~\ref{sec:BBK}); summaries of the conclusions for each model can be found at the end of their respective subsection. We conclude in section \ref{sec:concl}.

\section{Staggered Inflation: Fields Becoming Obsolete during Inflation \label{sec:main}}

\subsection{Background \label{sec:background}}

Consider $\mathcal{N}$ scalar fields with canonical kinetic terms and a separable potential $W=\sum_{A=1}^{\mathcal{N}}V_A$ \footnote{Note that $W$ is not the SUSY superpotential.}, so that the action reads
\begin{eqnarray}
S&=&\frac{1}{2}\int d^4x \sqrt{-g} \left(\frac{1}{2}\sum_{A=1}^{\mathcal{N}}\partial^\mu\varphi_A\partial_\mu\varphi_A+W(\varphi_1,\varphi_2,...)\right)\,.
\end{eqnarray}
For ease of notation, we set the reduced Planck mass equal to one throughout $m_p^2\equiv 1$. 
Assume that the fields all evolve according to the same potential $V_A \equiv V$ and that they start
out from an identical initial value \footnote{These assumptions simplify our treatment considerably, but are not crucial and could be relaxed while retaining the effect of staggered inflation.}, so that 
\begin{eqnarray}
W=\mathcal{N} V\,.
\end{eqnarray}
In this case, an effective single field model with $\varphi\equiv \sqrt{\mathcal{N}} \varphi_A$ and $W(\varphi)=\mathcal{N} V(\varphi/\sqrt{\mathcal{N}})$ describes correctly the inflationary phase (if $V$ is sufficiently flat) as well as the production of adiabatic perturbations (we comment on entropy perturbations in section \ref{sec:entropy}). To guarantee slow roll inflation, we demand
\begin{eqnarray}
\varepsilon_A\equiv\frac{1}{2}\left(\frac{V_A^\prime}{W}\right)^2\ll 1\,\,\,&,&\,\,\,   \varepsilon\equiv\frac{1}{2}\left(\frac{W^\prime}{W}\right)^2\ll 1\label{defepsilon} \,, \\
\eta_A\equiv \frac{V_A^{\prime\prime}}{W} \,\,\,&,&\,\,\,  |\eta_A|\ll 1 \,,\\
 \eta\equiv \frac{W^{\prime\prime}}{W} \,\,\,&,&\,\,\, |\eta| \ll 1\,,
\end{eqnarray}
where a prime on $V_A$ or $W$ denotes a derivative with respect to $\varphi_A$ or $\varphi$ respectively. So far, this is merely a simple model of assisted inflation \cite{Liddle:1998jc}, see  also \cite{Malik:1998gy,Kanti:1999vt} and follow ups.

Given this setup, we would like to investigate the consequences of individual fields $\varphi_A$ dropping out of the model, more or less instantaneously. By dropping out we mean that a field decays while its energy is converted into a different form, for example radiation. This may seem artificial at first glance, but it is actually quite generic in multi-field models of inflation within string theory \footnote{See \cite{Chialva:2008zw} for a related application within chain inflation.}. For instance, in the model of \cite{Becker:2005sg}  (see also \cite{Krause:2007jr,Ashoorioon:2006wc}) the inflatons are related to the distances between adjacent M5-branes. These branes are located along an orbifold $\bf{S}^1/\bf{Z}_2$ and slowly separate from each other in the orbifold direction, corresponding to the inflationary phase in the effective four dimensional description. Since the orbifold is quasi-static during this regime, the outermost branes will collide with the orbifold-fixed planes at some point in time, dissolving into the boundaries through small instanton transitions. Thus, the degrees of freedom associated with the distances to these just dissolved branes become obsolete during inflation. Of course, the energy associated with the branes does not vanish, but gets converted to other degrees of freedom, such as radiation. Another example is inflation driven by multiple tachyons as proposed in \cite{Majumdar:2003kd}; here, tachyons roll slowly away from the top of their potential, but fluctuations may displace a given field far enough to condense suddenly, again making obsolete this degree of freedom during inflation. We come back to these two concrete models later on, after developing the formalism to deal with decaying fields. For additional multi-field models see e.g. \cite{Cline:2005ty,Ward:2007gs,Grimm:2007hs,Dimopoulos:2005ac,Panigrahi:2007sq,Piao:2002vf,Mazumdar:2001mm} and \cite{Wands:2007bd} for a review.  

It is worth stressing at this point that the disappearance of a given field in the models considered in this paper is a sudden, but continuous process. This means potential problems associated with first order phase transitions, as discussed in some detail in \cite{Watson:2006px}, are absent. To model the disappearance of individual fields we promote $\mathcal{N}$ to a time dependent function. We further smooth out \footnote{It should be noted that by smoothing out $\mathcal{N}$, we will not be able to recover additional features in the power-spectrum that are directly related to  sudden drops in the potential. For instance, features could consist of a ringing or additional  non-Gaussianities \cite{Covi:2006ci,Ashoorioon:2006wc,Chen:2006xjb}. However, such signals depend crucially on the detailed physics of the fields' dissapearance, a rather badly understood and model dependent issue for the current generation of setups. As a consequence, these features will be quite hard to estimate within an effective single field description and we will not address them further in this article.} $\mathcal{N}(t)$ so that we can introduce a continuous decay rate $\Gamma\equiv -\dot{\mathcal{N}}/\mathcal{N}>0$, which is to be determined from the underlying model. This smoothing is the key simplifying assumption in our approach, and it is only viable if the number of fields is large and the decay rate is such that within any given Hubble time a few fields become obsolete.
For the cases we are interested in, this rate is small compared to the Hubble parameter $H$ during inflation (inflation ends quickly otherwise). This means we can introduce the small parameter
\begin{eqnarray}
\varepsilon_{\mathcal{N}}\equiv -\frac{\dot{\mathcal{N}}}{\mathcal{N}}\frac{1}{2H}=\frac{\Gamma}{2H}\sim \mathcal{O}(\varepsilon)\,. \label{defepsilonn}
\end{eqnarray}

The time dependence of $\mathcal{N}$ induces an additional decrease in the energy of the effective inflaton. To be concrete, the continuity equation of $\varphi$ needs to be modified to account for the energy loss in the inflaton sector due to decaying fields $\nabla_\mu T^{\mu 0}_\varphi = \dot{\mathcal{N}} V$, so that
\begin{eqnarray}
\dot{\rho}_\varphi=-3H(\rho_\varphi+p_\varphi)+\dot{\mathcal{N}}V\,.
\end{eqnarray}
Here and throughout the subscript $\varphi$ denotes the entire inflaton sector. It is worthwhile mentioning that the individual $\varphi_A$ evolve according to the standard slow roll equation of motion $3H\dot{\varphi}_A\simeq -\partial V_A /\partial \varphi_A$, as long as they are present. To retain  $\nabla_\mu T^{\mu 0}_{total} = 0$, we must allow for an additional component $\rho_r$ within the energy budget \footnote{Note that $\dot{N}<0$, so that energy always flows from the inflaton field sector into $\rho_{r}$.} satisfying
\begin{eqnarray}
\dot{\rho}_r=-3H(\rho_r+p_r)-\dot{\mathcal{N}}V\,. \label{energy_density}
\end{eqnarray}
We use the subscript ``$r$'', since we have radiation in mind ($p_r=w_r \rho_r$ with $w_r=1/3$); this choice seems most natural to us, considering the models we are interested in, but to remain as general as possible we keep $w_r$ arbitrary throughout.

The condition $\varepsilon_{\mathcal{N}}\ll 1$ implies that $\rho_r$  makes up only a small fraction of the total energy density during inflation. Thus, we may introduce the small parameter
\begin{eqnarray}
\bar{\varepsilon}&\equiv& \frac{3}{2}(1+w_r)\frac{\rho_r}{\rho_\varphi+\rho_r}\label{defbarepsilon}\\
&\sim& \mathcal{O}(\varepsilon)\,.
\end{eqnarray}
Naturally, $\bar{\varepsilon}$ is not independent of $\varepsilon_{\mathcal{N}}$, and one can show under mild assumptions that $\bar{\varepsilon}\rightarrow \varepsilon_{\mathcal{N}}$ during inflation (see below). Further, using slow roll of the individual fields $\varphi_A$, one can show that the total energy and pressure of the effective inflaton obey $\rho_\varphi+p_\varphi\simeq 2\varepsilon\rho_\varphi /3 $. This, along with $p_r=w_r \rho_r$ and the definitions above, leads to
\begin{eqnarray}
\dot{\rho}_\varphi&\simeq& -2H(\varepsilon_{\mathcal{N}}+\varepsilon)\rho_\varphi\,, \label{cont1}\\
\dot{\rho}_r&\simeq& -2H\left(\frac{3}{2}(1+w_r)\rho_r- \varepsilon_{\mathcal{N}}\rho_\varphi\right)\simeq2H (\varepsilon_{\mathcal{N}}-\bar{\varepsilon})\rho_{\varphi} \label{cont2}
\end{eqnarray}
during slow roll (the ``$\simeq$'' always denotes equality to first order in small parameters such as $\varepsilon, \eta, \varepsilon_{\mathcal{N}}$ or $\bar{\varepsilon}$). Taking the derivative with respect to time of the Friedmann equation $3H^2=\rho_\varphi+\rho_r$ and using (\ref{cont1}) as well as (\ref{cont2}), yields the {\it Hubble slow roll parameter}
\begin{eqnarray}
\hat{\varepsilon}&\equiv& -\frac{\dot{H}}{H^2} \label{defhatepsilon0}\\
&\simeq &\varepsilon+\bar{\varepsilon}\,. \label{defhatepsilon}
\end{eqnarray}
This explains the chosen pre-factors in (\ref{defepsilonn}) and (\ref{defbarepsilon}), as well as our demand that all epsilons should be small.

We have carefully avoided the use of the Klein Gordon equation for $\varphi$, since it gets modified by the presence of $\Gamma\neq 0$. To derive this modification, one can use $\rho_\varphi=\dot{\varphi}^2/2+W$ in the continuity equation (\ref{cont1}) and make the usual slow roll approximations,
or take the time derivative of $\varphi\equiv \sqrt{\mathcal{N}}\varphi_A$ directly with $3H\dot{\varphi}_A\simeq-\partial V_A/\partial \varphi_A$ so that
\begin{eqnarray}
3H\dot{\varphi}&\simeq& -W^\prime\gamma \label{eomsrvarphi}\,,
\end{eqnarray}
where we introduced the short hand notation
\begin{eqnarray}
\gamma&\equiv& 1+\varepsilon_{\mathcal{N}}\varphi \frac{W}{W^\prime}\,. \label{defgamma}
\end{eqnarray}
As expected, the usual slow roll equation of motion is recovered from (\ref{eomsrvarphi}) in the limit $\varepsilon_{\mathcal{N}}\rightarrow 0$.
The limit $W^\prime\rightarrow 0$, that is  $\varepsilon \rightarrow 0$, corresponds to a dynamically relaxing cosmological constant, which is discussed in great detail in \cite{Watson:2006px}.

During inflation, $H$, $\varepsilon_{\mathcal{N}}$ as well as $\rho_\varphi$ change very slowly, so that (\ref{cont2}) can be integrated to $\rho_r\simeq \varepsilon_{\mathcal{N}} 2\rho_\varphi/(3+3w_r)+ C \exp(-3(1+w_r)Ht)$ with $C=const$. This means that the additional component of the energy budget approaches a scaling solution $\rho_r\rightarrow  \varepsilon_{\mathcal{N}} 2\rho_\varphi/(3+3w_r)$ for which $\varepsilon_{\mathcal{N}}\simeq \bar{\varepsilon}$ and $\dot{\rho}_r\sim \mathcal{O}(\varepsilon_{\mathcal{N}})\dot{\rho}_\varphi\sim \mathcal{O}(\varepsilon_{\mathcal{N}}^2)\rho_\varphi \sim  \mathcal{O}(\varepsilon_{\mathcal{N}})\rho_r$. Thus, we can use
\begin{eqnarray}
\varepsilon_{\mathcal{N}}=\bar{\varepsilon}+\mathcal{O}(\varepsilon^2) \label{barepsilonandepsilonn}
\end{eqnarray}
 during inflation.

\subsubsection{Comparison to Warm Inflation \label{sec:warm}}
At this point, we would like to comment on similarities to warm inflation \cite{Berera:1995ie}, where the scalar field's interaction with other particles (through which the scalar field transfers some of its energy into a thermal bath) prevents the temperature from rapidly reaching zero. Within warm inflation, the motion of the scalar field is described by the modified Klein-Gordon equation (assuming slow roll)
\begin{eqnarray}
\dot\varphi\simeq -W'\frac{1}{3H+\tilde{\Gamma}}\,.\label{EOM_wI}
\end{eqnarray}
The extra friction term $\propto\tilde{\Gamma}$ represents an additional energy loss of the scalar field $\varphi$ stemming from particle creation.
This equation needs to be compared to our equation (\ref{eomsrvarphi}): in our case, an increase/decrease of the right hand side is present, depending on the sign of $\varphi/W^\prime$,
while in (\ref{EOM_wI}), the right hand side is always decreased by $\tilde{\Gamma}>0$.

 A more concrete realization of warm inflation based on thermal viscosity has been (critically) examined in \cite{Yokoyama:1998ju}. In this study, part of the inflaton's energy is converted into radiation through a viscosity term. As a consequence, one can show that the energy density satisfies
 \begin{eqnarray}
 \dot\rho_r=-4H\rho_r+C_v\dot\varphi^2\,,\label{energy_density_WI}
 \end{eqnarray}
with $C_v\gg 3H$. Comparing (\ref{energy_density_WI}) to our case (\ref{energy_density}) we see a similar modification: an additional positive term, potentially counterbalancing the dilution of $\rho_r$ due to redshifting; however, the origin of this term differs: in our case it is due to the transfer of potential energy, whereas that of (\ref{energy_density_WI}) arises from infusing kinetic energy.

Though the friction term in (\ref{energy_density_WI}) could at first glance be large enough to allow for successful warm inflation, a more careful examination by Yokoyama and Linde revealed \cite{Yokoyama:1998ju} that  warm inflation is not feasible in this framework, since $\varphi$ changes significantly over the relaxation time of the relevant particles, violating an adiabaticity condition required for the validity of (\ref{energy_density_WI}). Nevertheless, albeit the presence of many problems like the one just mentioned, warm inflation remains an active field of research.

Staggered multi-field inflation, as introduced in the present paper, might be seen as an independent (less problematic) realization of warm inflation, which can be embedded into string theory.

\subsection{The Scalar Spectral Index \label{sec:spectralindex}}
In order to compute the scalar spectral index we focus on adiabatic perturbations, that is, for the time being we neglect isocurvature perturbations. The latter ones arise due to the presence of the many scalar fields and $\rho_r$, but one can show that they are suppressed during inflation. The applicability of this approximation is discussed in section \ref{sec:entropy}.

The canonical degree of freedom that diagonalizes the action of adiabatic scalar perturbations is the Mukhanov variable $v_k$ \cite{Mukhanov:1990me}. It is related to the curvature perturbation on uniform density hyper-surfaces $\zeta_k$ via
\begin{eqnarray}
v_k=z\zeta_k\,,
\end{eqnarray}
where
\begin{eqnarray}
z&\equiv&\frac{1}{\theta c_s}\,,\\
\theta^2&\equiv& \frac{1}{3a^2(1+w)}\,.\label{deftheta}
\end{eqnarray}
Here, $w=p/\rho$ with $\rho=\rho_\varphi+\rho_r$ and $p=p_\varphi+p_r$ is the equation of state parameter, while  $c_s^2=\partial p/ \partial \rho|_{S=const}$ is the adiabatic sound speed, which becomes  $c_s^2\approx \dot{p}/\dot{\rho}$ on large scales. In terms of $\zeta$, the power spectrum is given by
\begin{eqnarray}
\mathcal{P}_\zeta=\frac{k^3}{2\pi^2}|\zeta_k|^2\,, \label{defpowerspectrum}
\end{eqnarray}
and the scalar spectral index is
\begin{eqnarray}
n_s-1=\frac{d\ln \mathcal{P}_\zeta}{d \ln k}\,.
\end{eqnarray}
The Mukhanov variable satisfies the simple equation of motion \cite{Mukhanov:1990me}
\begin{eqnarray}
v_k^{\prime\prime}+\left(k^2c_s^2-\frac{z^{\prime\prime}}{z}\right)v_k=0\,, \label{eomvk}
\end{eqnarray}
where a prime denotes a derivative with respect to conformal time $\tau=-\infty\dots0$, $a\, d\tau=dt$. If the Hubble slow roll parameter $\hat{\varepsilon}$ in (\ref{defhatepsilon0}) is evolving slowly, which is the case during inflation, we can approximate
\begin{eqnarray}
a(\tau)\propto (-\tau)^{-(1+\hat{\varepsilon})}\,, \label{aoftau}
\end{eqnarray}
and solve (\ref{eomvk}) analytically in terms of Hankel functions (see e.g. \cite{Mukhanov:1990me} or \cite{Watson:2006px}).
Writing
\begin{eqnarray}
\frac{z^{\prime\prime}}{z}\equiv \frac{\nu^2-1/4}{\tau^2}\,,
\end{eqnarray}
treating $\nu$ as a constant and imposing the Bunch-Davies vacuum  at $kc_s\tau\rightarrow -\infty$  ($v_k=\exp(-ikc_s\tau)/\sqrt{2c_sk}$), we get
\begin{eqnarray}
v_k=\frac{M}{2}\sqrt{-\tau\pi}H_\nu^{(1)}(-kc_s\tau)\,,
\end{eqnarray}
where $M$ is an irrelevant phase factor, $|M|=1$. Hence, the curvature perturbation reads
\begin{eqnarray}
|\zeta_k|=\frac{1}{2z}\sqrt{-\tau\pi}H_\nu^{(1)}(-kc_s\tau)\,,
\end{eqnarray}
which can be expanded on large scales to
\begin{eqnarray}
|\zeta_k|\approx \frac{1}{z\sqrt{2 k c_s}}(-kc_s\tau)^{1/2-\nu}\,.
\end{eqnarray}
Plugging this into (\ref{defpowerspectrum}) and taking the logarithmic derivative, one can read off the scalar spectral index to
\begin{eqnarray}
n_s-1=3-2\nu\,.
\end{eqnarray}

Thus, we only have to compute $z^{\prime\prime}/z$ and identify $\nu$ in order to get $n_s$. Since $c_s^{\prime}$ is second order in small parameters, we need to focus on $\theta$ from (\ref{deftheta}) only. To leading order in $\varepsilon$ and $\varepsilon_\mathcal{N}$, the equation of state parameter reads

\begin{eqnarray}
w\simeq-1+\frac{2}{3}(\varepsilon\gamma^2+\varepsilon_{\mathcal{N}})\,.
\end{eqnarray}
Here, we used the equation of motion for $\varphi$ in (\ref{eomsrvarphi}), $\gamma$ from (\ref{defgamma}) and $\bar{\varepsilon}\simeq \varepsilon_{\mathcal{N}}$ from (\ref{barepsilonandepsilonn}). Consequently
\begin{eqnarray}
\frac{1}{\theta^2}\simeq a^22(\varepsilon\gamma^2+\varepsilon_{\mathcal{N}})\,.
\end{eqnarray}
In order to take the derivatives with respect to conformal time, we need
\begin{eqnarray}
\varepsilon^\prime&\simeq&aH\gamma(4\varepsilon^2-2\varepsilon\eta)\,, \label{derep}\\
\varepsilon_{\mathcal{N}}^\prime&\simeq&aH\varepsilon_{\mathcal{N}}(\varepsilon+\varepsilon_{\mathcal{N}})(1-\delta)\,,\\
\gamma^\prime&\simeq&aH\left[(\gamma-1)\left[(\varepsilon+\varepsilon_{\mathcal{N}})(1-\delta)+\gamma^3(\eta-2\varepsilon)\right]-\varepsilon_{\mathcal{N}}\gamma\right]\,, \label{dergam}
\end{eqnarray}
where we introduced
\begin{eqnarray}
\delta\equiv \frac{\dot{\Gamma}H}{\Gamma\dot{H}}\,.
\end{eqnarray}
Note that $\delta=1$ corresponds to $\varepsilon_{\mathcal{N}}=const$, whereas $\delta=0$ corresponds to $\Gamma=const$. Deriving (\ref{derep})-(\ref{dergam}) based on section \ref{sec:background} is straightforward, albeit somewhat tedious. After some more algebra, we arrive at
\begin{eqnarray}
\nonumber \frac{z^{\prime\prime}}{z}&\simeq& 2a^2H^2\bigg[1-\frac{1}{2}(\varepsilon+\varepsilon_{\mathcal{N}})+\frac{3}{2}\frac{1}{\varepsilon\gamma^2+\varepsilon_{\mathcal{N}}}\Big(
\gamma^3\left(2\varepsilon^2-\varepsilon\eta\right)\\
&&+\varepsilon\gamma\left[(\gamma-1)\left[(\varepsilon+\varepsilon_{\mathcal{N}})(1-\delta)+\gamma(\eta-2\varepsilon)\right]-\varepsilon_{\mathcal{N}}\gamma\right]
+\frac{1}{2}\varepsilon_{\mathcal{N}}(\varepsilon+\varepsilon_{\mathcal{N}})(1-\delta)
\Big)\bigg]\,,
\end{eqnarray}
to leading order in small parameters (we assume that $\delta^\prime$ is of the same order as $\varepsilon_{\mathcal{N}}$).
Using $a(\tau)$ from (\ref{aoftau}) so that $a^2H^2\simeq (1+2\varepsilon+2\varepsilon_{\mathcal{N}})/\tau^2$ we can read off $\nu$. The resulting scalar spectral index is
\begin{eqnarray}
n_s-1&\simeq&-2(\varepsilon+\varepsilon_{\mathcal{N}})-\frac{2}{\varepsilon\gamma^2+\varepsilon_{\mathcal{N}}}\bigg[\varepsilon\gamma^2(2\varepsilon-\varepsilon_{\mathcal{N}}-\eta)+(\varepsilon+\varepsilon_{\mathcal{N}})(1-\delta)(\varepsilon\gamma(\gamma-1)+\frac{\varepsilon_{\mathcal{N}}}{2})\bigg].\label{finalns}
%
%
%
\end{eqnarray}
This is our first major result. In the limit $\Gamma=0$, so that $\varepsilon_{\mathcal{N}}=0$ and $\gamma=1$, we recover the slow roll result $n_s^{SR}-1\simeq -6\varepsilon+2\eta$. On the other hand, if $\varepsilon_{\mathcal{N}}=const$ (that is $\delta=1$) and $\varepsilon=\eta=0$, we recover the case of a dynamically relaxing cosmological constant of \cite{Watson:2006px} $n_s^{relax.\,CC}-1=-2\varepsilon_{\mathcal{N}}$.

\subsection{Running \label{sec:running}}
The running $\partial n_s/\partial \ln k$ can be computed by applying\footnote{At horizon crossing $k=aH$, resulting in $d \ln k\simeq aH d\tau$; we use that $H$ is evolving slowly during inflation, that is $-\dot{H}/H^2=\hat{\varepsilon} \simeq \varepsilon + \varepsilon_{\mathcal{N}} \ll 1$.} $\partial/\partial \ln k=(aH)^{-1}\partial/\partial \tau$  to  (\ref{finalns}). Using again $\varepsilon_{\mathcal{N}}\simeq \bar{\varepsilon}$ and
\begin{eqnarray}
\eta^\prime \simeq -aH\gamma \left(-2\varepsilon\eta+ \xi^2\right)\,,
\end{eqnarray}
where $\xi^2\equiv W^\prime W^{\prime\prime\prime}/W^2$, as well as $\varepsilon^\prime$, $\varepsilon_{\mathcal{N}}^{\prime}$ and $\gamma^\prime$ from (\ref{derep})-(\ref{dergam}) the running reads
\begin{eqnarray}
\label{finalrunning} aH\frac{\partial n_s}{\partial \ln k}&\simeq& -2(\varepsilon'+\varepsilon_{\mathcal N}')-\frac{2}{\varepsilon\gamma^2+\varepsilon_{\mathcal N}}\Bigg[(\varepsilon'\gamma^2+2\varepsilon\gamma\gamma')(2\varepsilon-\eta-\varepsilon_{\mathcal N})+\varepsilon\gamma^2(2\varepsilon'-\eta'-\varepsilon_{\mathcal N}')\\
&&+\nonumber(\varepsilon'+\varepsilon_{\mathcal N}')(1-\delta)\left[\varepsilon\gamma(\gamma-1)+\frac{\varepsilon_{\mathcal N}}{2}\right]\\
&&+\nonumber(\varepsilon+\varepsilon_{\mathcal N})\left[(1-\delta)\left[(\gamma-1)(\varepsilon'\gamma+\varepsilon\gamma')+\varepsilon\gamma\gamma'+\frac{\varepsilon_{\mathcal N}'}{2}\right]-\delta'\left[\varepsilon\gamma(\gamma-1)+\frac{\varepsilon_{\mathcal N}}{2}\right]\right]\Bigg]\\
&&-\nonumber\frac{2[\varepsilon'\gamma^2+2\varepsilon\gamma\gamma'+\varepsilon_{\mathcal N}']}{(\varepsilon\gamma^2+\varepsilon_{\mathcal N})^2}\left[\varepsilon\gamma^2(\eta+\varepsilon_{\mathcal N}-2\varepsilon)-(\varepsilon+\varepsilon_{\mathcal N})(1-\delta)\left[\varepsilon\gamma(\gamma-1)+\frac{\varepsilon_{\mathcal N}}{2}\right]\right]\,.
\end{eqnarray}
In the limit $\varepsilon_{\mathcal{N}}\rightarrow 0$, we recover the standard slow roll result\footnote{Note a recurring sign mistake in the literature, e.g. in the popular review \cite{Lyth:1998xn} or textbook \cite{Liddle:2000cg}; see \cite{Kosowsky:1995aa} for the correct expression.} $\partial n_s/\partial \ln k\simeq 16\varepsilon \eta-24 \varepsilon^2-2\xi^2 $. On the other hand, if $\varepsilon=0$, the running reduces to $\partial n_s/ \partial \ln k=-\varepsilon_{\mathcal{N}}^2(3-4\delta+\delta^2)+\varepsilon_{\mathcal{N}}\delta^{\prime}$. In any event, the running remains second order in small parameters, rendering it un-observably small.

\subsection{Isocurvature (Entropy) Perturbations \label{sec:entropy}}
So far we focused on adiabatic perturbations only, neglecting entropy perturbations entirely. To check if entropy perturbations are indeed negligible, we follow  \cite{Malik:2002jb} (see also \cite{Kodama:1985bj,Watson:2006px}). The perturbation of the total pressure is in general
\begin{eqnarray}
\delta p =c_s^2 \delta \rho + \tau \delta S\,,
\end{eqnarray}
where $S$ is the entropy density, $\tau = \partial p / \partial s|_\rho$ (we use cosmic time in this section in order to avoid confusion of this $\tau$ with conformal time) and $c_s^2 = \partial p / \partial \rho|_s$ is the adiabatic sound speed, which reduces to $c_s^2= \dot{p}/\dot{\rho}$ on large scales. The non-adiabatic pressure perturbation may thus be defined as
\begin{eqnarray}
\delta p_{nad}\equiv \tau \delta S\,.
\end{eqnarray}
In a multi-component fluid, there are two contributions to $\delta p_{nad}$,  a relative one $\delta p_{nad}^{rel}$ between the fluid components and an intrinsic one within each fluid $\delta p_{nad}^{int}$ \cite{Malik:2002jb}. In our case, we have two main components, the effective inflaton (composed of $\mathcal{N}$ fields) with $\rho_\varphi$, and the additional component $\rho_r$, for instance given by radiation, with a constant equation of state parameter $w_r=\rho_r/p_r$.

Let us first discuss $p_{nad}^{rel}$, which can be written as \cite{Malik:2002jb}
\begin{eqnarray}
\delta p_{nad}^{rel}=\frac{1}{3H\dot{\rho}}\dot{\rho}_r\dot{\rho}_\varphi(c_r^2-c_\varphi^2)S_{r\varphi}
\end{eqnarray}
where $c_\alpha^2=\dot{p}_\alpha/\dot{\rho}_\alpha$ and the relative entropy perturbation is defined as
\begin{eqnarray}
S_{r\varphi}&=&3(\zeta_r-\zeta_\varphi)\\
&=&-3H\left(\frac{\delta \rho_r}{\dot{\rho}_r}-\frac{\delta \rho_\varphi}{\dot{\rho}_\varphi}\right)\,.
\end{eqnarray}
We have already seen in the discussion before (\ref{barepsilonandepsilonn}) that $\rho_r$ approaches a scaling solution $\rho_r\rightarrow  \varepsilon_{\mathcal{N}} 2\rho_\varphi/(3+3w_r)$ during inflation for which $\dot{\rho}_r$ is small ($\dot{\rho}_r\sim \mathcal{O}(\varepsilon_{\mathcal{N}})\dot{\rho}_\varphi\sim \mathcal{O}(\varepsilon_{\mathcal{N}}^2)\rho_\varphi \sim  \mathcal{O}(\varepsilon_{\mathcal{N}})\rho_r$). Since  $\delta p_{nad}^{rel} \propto  \dot{\rho}_r$, the relative non-adiabatic pressure perturbation is heavily suppressed and can safely be ignored during inflation \footnote{Note that $S_{r\varphi}$ remains small and finite in the limit of small $\dot{\rho}_r$, because the curvature perturbations $\zeta_r$ and $\zeta_{\varphi}$ remain small; see \cite{Watson:2006px}.}.

Second, consider the intrinsic contributions
\begin{eqnarray}
\delta p_{nad}^{int}&=&\delta p_{nad,r}^{int}+\delta p_{nad,\varphi}^{int}\\
&=& \delta p_r-c_r^2\delta \rho_r+\delta p_\varphi-c_\varphi^2\delta\rho_\varphi\,.
\end{eqnarray}
Since $w_r=const$ we have $w_r=c_r^2$ and $\delta p_r= w_r \delta \rho_r$. Therefore $\delta p_{nad,r}^{int}=\delta p_r-c_r^2\delta \rho_r=\delta \rho_r(c_r^2-c_r^2)=0$,
that is the intrinsic non-adiabatic pressure $\delta p _{nad,r}^{int}$ vanishes identically. However, for the effective inflaton the case is less clear, since its equation of state (and thus the intrinsic sound speed) changes and $\varphi$ is actually composed of $\mathcal{N}$ components. We first note that the individual fields approach an equation of state  $p_{\varphi_A}\simeq -\rho_{\varphi_A}$ during inflation, so that the equation of state parameter becomes nearly constant for each of the fields. Thus we have $\delta p_{\varphi_A}\simeq w_{\varphi_A} \delta \rho_{\varphi{A}}$ where $w_{\varphi_A}\simeq c_{\varphi_A}^2\simeq -1$. Therefore, the intrinsic non-adiabatic pressure within each inflaton field is $\delta p_{nad,\varphi_A}^{int}=\delta p_{\varphi_A}-c_{\varphi_A}^2\delta \rho_{\varphi_A}\simeq \delta \rho_{\varphi_A}(c_{\varphi_A}^2-c_{\varphi_A}^2)=0$. Second, the $\mathcal{N}$ components' relative non-adiabatic pressure perturbations contribute to $\delta p_{nad,\varphi}^{int}$. But if we look at the relative non-adiabatic pressure contributions between the fields $\delta p_{nad,\varphi_A\varphi_B}^{rel}\propto (c^2_{\varphi_A}-c^2_{\varphi_B})$ we conclude that $\delta p_{nad,\varphi_A\varphi_B}^{rel}\simeq 0$ since $c^2_{\varphi_A}\simeq c^2_{\varphi_B} \simeq -1$. Thus, the total non-adiabatic pressure is negligible during inflation and we are justified to focus on adiabatic perturbations \footnote{There is one caveat to the above arguments: during the short intervals when fields decay the evolution of the decaying field is rapid and its equation of state changes -- it would indeed be interesting to investigate the productions of isocurvature perturbations (and non-Gaussianities) during these instances, which can not be recovered with our approach since we employ a smooth $\mathcal{N}(t)$ \cite{Ashoorioon}.}. These are correctly recovered by focusing on the Mukhanov variable $v_k$, as we did in section \ref{sec:spectralindex}.

One further comment might be in order: due to the suppression of entropy perturbations during inflation, we do not expect any large additional non-Gaussianities (NG) caused by $\Gamma\neq 0$. Further, since multi-field inflationary models generically yield comparable NG to their single field analogs \cite{Battefeld:2006sz,Battefeld:2007en}, and NG are suppressed during slow roll inflation (see however \cite{Byrnes:2008wi}), we do not expect any measurable NG within the setups discussed in this paper. One caveat to this argument consists of the short intervals whenever one of the inflaton fields decays (see e.g. \cite{Chen:2006xjb} for NG from steps in a potential in single field inflation). We cannot exclude the production of NG at these instances, but we do not anticipate them either, because the process is very similar to preheating and generically, preheating (e.g. instant preheating) does not cause large NG  \cite{Enqvist:2005qu} (see however the possibility of larger NG in tachyonic preheating \cite{Enqvist:2005qu,Barnaby:2006cq,Barnaby:2006km}). Nevertheless, given a better microphysical understanding of how inflatons become obsolete, e.g. by investigating the brane annihilation in \cite{Majumdar:2003kd,Becker:2005sg}, one can and should check the validity of this expectation.

\section{Applications \label{sec:models}}
We would like to compute the scalar spectral index in (\ref{finalns}) and the running in (\ref{finalrunning}) within a couple of models. First, we extend assisted inflation with exponential potentials \cite{Liddle:1998jc} by incorporating a non-zero $\Gamma$, which provides a graceful exit of inflation; this phenomenological model has the advantage of being instructive and simple. Next, we consider two concrete models \cite{Majumdar:2003kd,Becker:2005sg} which have the feature of decaying fields during inflation already build in, at the price of being more complicated to treat. Our approach consists of extracting the potential slow roll parameters $\varepsilon$, $\eta$ and $\xi$, as well as $\varepsilon_{\mathcal{N}}=\Gamma/(2H)$ with $\Gamma=-\dot{\mathcal{N}}/\mathcal{N}$ in order to apply (\ref{finalns}) and (\ref{finalrunning}).

\subsection{Staggered Assisted Inflation \label{sec:staggered}}
Consider the original proposal of assisted inflation \cite{Liddle:1998jc}, where the $\mathcal{N}$ scalar fields have identical exponential potentials
\begin{eqnarray}
V_A=V_0e^{-\sqrt{\frac{2}{p}}\varphi_A}\,,
\end{eqnarray}
so that the potential for the single effective field $\varphi=\sqrt{\mathcal{N}}\varphi_A$ reads
\begin{eqnarray}
W=\mathcal{N}V_0e^{-\sqrt{\frac{2}{\tilde{p}}}\varphi}\,,\label{defWstaggered}
\end{eqnarray}
with $\tilde{p}=\mathcal{N}p$ (we assume identical initial values for all $\varphi_A$). Power law inflation ($a\propto t^{\tilde{p}}$) results for large enough $\tilde{p}$, which can be achieved even with steep potentials if $\mathcal{N}\gg 1$. Note that the single field solution is an attractor during inflation \cite{Malik:1998gy} (see \cite{Calcagni:2007sb} for a general discussion of stability in multi-field inflation).
 The slow roll parameters in the above model are
\begin{eqnarray}
\eta=2\varepsilon=\frac{2}{\tilde{p}}\,, \label{SRparametersassisted}
\end{eqnarray}
so that the scalar spectral index is $n_s-1=-2/\tilde{p}$ if $\Gamma=0$ (see \cite{Liddle:1998jc} or equation (\ref{finalns})); in addition, the running is zero since $n_s=const$.

However, the above model has a graceful exit problem: inflation never ends because $\varepsilon$ and $\eta$ are constant. This problem can be alleviated by introducing a non-zero $\Gamma$, so that the number of fields decreases during inflation. Let's for simplicity take a constant rate $\Gamma=const$ so that $\delta=0$. Consequently, inflation comes to an end when $\varepsilon_{\mathcal{N}}=\Gamma/2H$ becomes of order one, at which point $\mathcal{N}$ decreases rapidly during a Hubble time so that the assistance effect diminishes.

Since a shift in the individual fields $\varphi_A\rightarrow \varphi_A+const$ can be absorbed into a redefinition of $V_0$, we can set $\varphi=0$ at $N=60$ efolds before the end of inflation, without loss of generality. Hence $\gamma=1$, and the scalar spectral index in (\ref{finalns}) becomes \footnote{A shift in the fields causes both, a shift in $\gamma$ since $\varphi$ changes, and  a shift in $\varepsilon_{\mathcal{N}}$ since $V_0$ changes; however, an observable such as the scalar spectral index remains unaffected.}
\begin{eqnarray}
n_s-1\simeq-2\varepsilon-3\varepsilon_{\mathcal{N}}+2\frac{\varepsilon\varepsilon_{\mathcal{N}}}{\varepsilon+\varepsilon_{\mathcal{N}}}\,, \label{assistedns}
\end{eqnarray}
where we used $\eta=2\varepsilon$ and $\delta=0$. Similarly, the running in (\ref{finalrunning}) reads
\begin{eqnarray}
\frac{\partial n_s}{\partial \ln k}&\simeq&-\frac{\varepsilon_{\mathcal{N}}}{(\varepsilon+\varepsilon_{\mathcal{N}})^2}\left(3\varepsilon_{\mathcal{N}}^3+13 \varepsilon\varepsilon_{\mathcal{N}}^2+5\varepsilon^2\varepsilon_{\mathcal{N}}-\varepsilon^3\right)\,,
\end{eqnarray}
where we also used $\xi^2=4\varepsilon^2$ and $\delta^\prime=0$.

These predictions differ from the corresponding slow roll ones: the additional energy loss in the inflaton sector due to $\Gamma\neq 0$ causes a redder spectrum with a running that is second order in the epsilons  (such a running is well below current observational limits \footnote{Since the running is generically second order in the slow roll parameters and $\varepsilon_{\mathcal{N}}$, we will not comment on it further in section \ref{sec:ACDMM} and \ref{sec:BBK}.}). The physical reason for the difference in the spectrum is the smooth graceful exit caused by a non-zero decay rate, resulting in $\varepsilon_\mathcal{N}\neq 0$. Note that for a sharp exit from inflation to the reheating era, which could be modeled by a step function $\Gamma=\Gamma_0 \theta(\varphi-\varphi_{end})$ with $\Gamma_0\gg 2H(\varphi_{end})$, we anticipate no corrections.

\begin{figure}[tb]
\includegraphics[scale=0.5,angle=0]{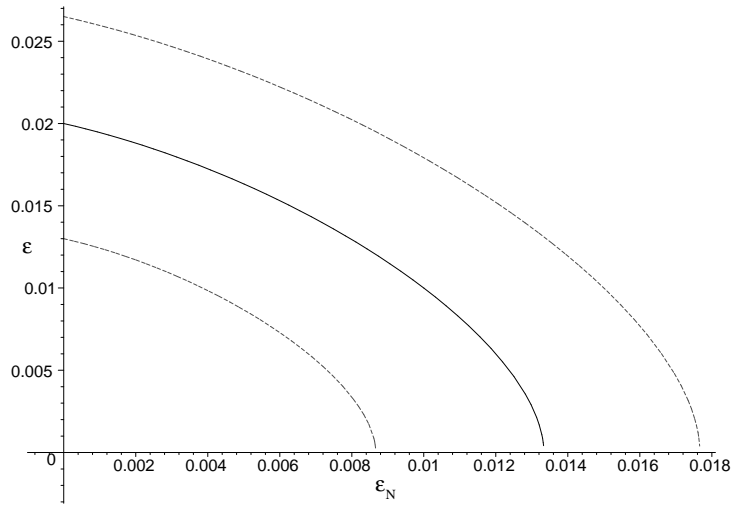}
   \caption{\label{pic:assistedns} Using $\eta=2\varepsilon$ (assisted inflation with exponential potentials) we plot $n_s-1=const$ lines in the $\varepsilon$-$\varepsilon_{\mathcal{N}}$ plane using (\ref{assistedns}), showing the $1\sigma$ interval of WMAP5 \cite{Komatsu:2008hk} $n_s=0.960^{+0.014}_{-0.013}$ (solid and dotted lines). Thus, a non-zero $\varepsilon_{\mathcal{N}}$ requires a lower value of $\varepsilon$ in order to remain consistent with observations. }
\end{figure}

In figure \ref{pic:assistedns}, we plot $n_s-1=const$ lines in the $\varepsilon$-$\varepsilon_{\mathcal{N}}$ plane, showing the $1\sigma$ interval of WMAP5 \cite{Komatsu:2008hk} $n_s=0.960^{+0.014}_{-0.013}$. Given $\varepsilon_{\mathcal{N}}>0$, it is evident that in order to fit $n_s$ from observations in assisted inflation we need a smaller $\varepsilon$. To put it another way, the smooth graceful exit of inflation introduced by $\Gamma\neq 0$ comes at the price of either requiring flatter potentials or more fields. Further,  the spectral index lies outside the $1\sigma$ region if $\varepsilon_{\mathcal{N}}$ is too large ($\varepsilon_{\mathcal{N}}=\Gamma/(2H)>0.018$), even if $\varepsilon$ were identical to zero. This means the half-time $t_\Gamma\equiv\ln(2)/\Gamma$ needs to be considerably larger than the Hubble time $t_H=H^{-1}$ during inflation, in order to be consistent with observations ($t_\Gamma\gtrsim 19\, t_H$).

In the phenomenological model above, we inserted a non-zero decay rate  by hand. As we shall see in the next sections, more concrete models within string theory actually force us to have $\Gamma\neq 0$, rendering the above model more natural.

\subsection{Inflation from Multiple Tachyons \label{sec:ACDMM}}
In \cite{Majumdar:2003kd}, $\mathcal{N}$ brane anti-brane pairs ($Dp-\bar{D}p$) where considered, giving rise to $\mathcal{N}^2$ tachyons due to a $U(\mathcal{N})\times U(\mathcal{N})$ gauge symmetry. However, these tachyons are generically all coupled to each other, making them not well suited for assisted inflation. To alleviate this problem, Davis and Majumdar proposed to focus on the Abelian part of $U(\mathcal{N})\times U(\mathcal{N})$, resulting in $\mathcal{N}$ uncoupled tachyons  $\varphi_A$ \footnote{Our notation differs from \cite{Majumdar:2003kd}, where tachyons are denoted by $t_i$ and time by $\tau$.}, $A=1\dots \mathcal{N}$; admittedly, this choice is to some extent  unphysical \cite{Majumdar:2003kd}, but offers an instructive model. Within this setup, the potential has the form \cite{Ohmori:2001am}
\begin{eqnarray}
W=\mathcal{N}\tau_p -c_1\sum_{A=1}^{\mathcal{N}}|\varphi_A|^2+c_2\sum_{A=1}^{\mathcal{N}}|\varphi_A|^4+\mathcal{O}\left(|\varphi_A|^6\right)\,, \label{potentialacd}
\end{eqnarray}
which is valid in close proximity to $(\varphi_1,...,\varphi_{\mathcal{N}})=0$. Here, $c_1\approx 0.87$ and $c_2\approx 0.21$ \cite{Ohmori:2001am}, but the brane tension $\tau_p$ is model dependent. For large $\mathcal{N}$ sufficient inflation can result if the tachyons are initially  close to zero. For simplicity, we assume that they all start out from the same initial value, in line with our assumptions in section \ref{sec:main}.

Whenever a tachyon gets displaced far enough from the origin (be it due to its slow roll evolution or a dislocation caused by either thermal or quantum mechanical fluctuations  \cite{Majumdar:2003kd}) the perturbative potential in (\ref{potentialacd}) becomes unreliable, and the tachyon condenses quickly. This condensation corresponds to the annihilation of a brane anti-brane pair. Inflation ends when all tachyons condense. This condensation is also expected to be responsible for reheating, but a concrete study is lacking in the literature. Giving the model the benefit of the doubt, we assume that whenever a brane anti-brane pair annihilates, its energy is indeed converted into some type of relativistic matter with $p_r=\rho_r/3$ (from the four dimensional point of view). Thus, the model of \cite{Majumdar:2003kd} is exactly of the type we examined in section \ref{sec:main}.

Next, we extract the slow roll parameters as well as specify the rate $\Gamma(t)=-\dot{\mathcal N}/\mathcal{N}$ at which the tachyons condense in order to compute $n_s$. If all tachyons condense at the same time, we have $\Gamma=0$ during inflation and the standard slow roll expressions apply. However, given the sensitivity of condensation to thermal and quantum mechanical dislocations of the tachyons, a staggered fashion of condensation is expected. In \cite{Majumdar:2003kd} three types of such a staggered condensation were proposed:
\begin{enumerate}
\item The number of tachyons decreases exponentially so that $\Gamma=const$ ($\Gamma = \ln 2/\tau_*$ in the notation of \cite{Majumdar:2003kd}), similar to the model in section \ref{sec:staggered}.
\item Tachyons condense serially so that $\mathcal{N}(t)=\mathcal{N}_0(1-t/\tilde{t})$ and $\Gamma =1/(\tilde{t}-t)$ ($\tilde{t}/\mathcal{N}_0=\tau_*$ in the notation of \cite{Majumdar:2003kd}). This is the case if tachyons condense at wildly different times.
\item Tachyons condense in a staggered fashion, but a handful survive and drive an extended phase of slow roll inflation, rendering again $\Gamma=0$ during the cosmological relevant phase of inflation. (We do not examine this case further.)
\end{enumerate}

Because the applicability of the potential (and thus slow roll) is questionable for large values of $\varphi$, we first examine cases one and two in a simplified setup with all tachyons sitting very close to the top of their potentials ($\varphi\approx 0$, $\dot{\varphi}\approx 0$). Whenever a tachyon gets displaced by a fluctuation, it is assumed to condense immediately. In later sections, we incorporate slow roll.

\subsubsection{$\Gamma\neq 0$, Negligible Slow Roll Contributions \label{sec:acdwithoutslowroll}}
{\bf Case 1) $\Gamma=const$}\\
Similar to section \ref{sec:staggered}, we consider a constant condensation rate so that the number of tachyons decreases exponentially in time
\begin{eqnarray}
\mathcal{N}(t)=\mathcal{N}_0 e^{-\Gamma t}\,,
\end{eqnarray}
where we take $t_{ini}\equiv 0$ at $N\approx 60$ efolds before the end of inflation. Once the number of fields is depleted, that is once $\mathcal{N}\sim 1$, inflation ends. To be concrete, we take $t_{end}=\ln(\mathcal{N}_0)/\Gamma$. Then, using $W(\varphi=0)=\mathcal{N}\tau_p$ and $H\approx \sqrt{W/3}$, the number of efolds becomes
\begin{eqnarray}
N&=&\int_{t_{ini}}^{t_{end}}H\, dt\\
&\approx& \frac{2}{\Gamma}\left(\frac{\tau_p\mathcal{N}_0}{3}\right)^{1/2}\left(1-\frac{1}{\sqrt{\mathcal{N}_0}}\right)\\
&\approx& \frac{2}{\Gamma}\left(\frac{\tau_p\mathcal{N}_0}{3}\right)^{1/2}\,. \label{efoldswsrconstGamma}
\end{eqnarray}
Further, since
\begin{eqnarray}
\varepsilon_{\mathcal{N}}=\frac{\Gamma}{2}\left(\frac{3}{\tau_p\mathcal{N}_0}\right)^{1/2}\approx \frac{1}{N}
\end{eqnarray}
we get from (\ref{finalns})
\begin{eqnarray}
n_s-1\simeq -3\varepsilon_{\mathcal{N}} \approx -\frac{3}{N} \label{ns3overN}
\end{eqnarray}
which is within the $1\sigma$ error bars of WMAP5 \cite{Komatsu:2008hk}.

If we fine-tune the brane tension such that $W$ vanishes at the minimum, $\tau_p\equiv c_1^2/(4c_2)\approx 0.90$, we obtain
\begin{eqnarray}
\frac{\mathcal{N}_0}{\Gamma^2}\approx \frac{3N^2c_2}{c_1^2}\approx 3000\,.
\end{eqnarray}
Thus, for $\Gamma\sim\mathcal{O}(1)$, we need $\mathcal{N}_0\sim 3000$ tachyons, which is somewhat large; however, for $\Gamma\lesssim 1$ we achieve  the desired amount of inflation and a spectral index within observational bounds with a few hundred tachyons.

It should be noted that neglecting slow roll might actually be the best motivated case: we only trust the tachyon potential close to $\varphi_A=0$; then the potential is indeed very flat so that $\varepsilon$ can be neglected \footnote{Note that in the present case the amplitude of perturbations (set by the COBE normalization) is determined by the decay rate, that is $\varepsilon_{\mathcal{N}}$ (just as in \cite{Watson:2006px}), and not by $\varepsilon$.}. As soon as an individual tachyon gets dislodged a bit, it should quickly condense and drop out of the model, leading to the above estimate for $n_s$.

{\bf Case 2) $\Gamma=(\tilde{t}-t)^{-1}$}\\
Here, the number of fields is decreasing linearly
\begin{eqnarray}
\mathcal{N}(t)=\mathcal{N}_0\left(1-\frac{t}{\tilde{t}}\right)
\end{eqnarray}
so that inflation ends around $t_{end}=\tilde{t}(1-1/\mathcal{N}_0)$. Analogous to (\ref{efoldswsrconstGamma}), the number of efolds becomes
\begin{eqnarray}
N&\approx& \frac{2\tilde{t}}{3}\left(\frac{\tau_p\mathcal{N}_0}{3}\right)^{1/2}\left(1-\frac{1}{\mathcal{N}_0^3}\right)\\
&\approx& \frac{2\tilde{t}}{3}\left(\frac{\tau_p\mathcal{N}_0}{3}\right)^{1/2}\,.
\end{eqnarray}
so that $\varepsilon_{\mathcal{N}}\approx 1/(3N)$. Since $\delta\simeq -2\varepsilon_{\mathcal{N}}/\hat{\varepsilon}\simeq -2$ we get from (\ref{finalns})
\begin{eqnarray}
n_s-1\simeq -5\varepsilon_{\mathcal{N}} \approx -\frac{5}{3N}\,,
\end{eqnarray}
 smaller than in the $\Gamma=const$ case and close to the $1\sigma$ boundary of WMAP5 \cite{Komatsu:2008hk}. Tuning the brane tension again to  $\tau_p\equiv c_1^2/(4c_2)\approx 0.90$, we obtain
\begin{eqnarray}
\mathcal{N}_0\tilde{t}^2\approx \frac{27N^2c_2}{c_1^2}\approx 27000\,.
\end{eqnarray}
As a result, in order to achieve the desired $N=60$ efolds of inflation with a few hundred fields we need $\tilde{t}\sim 10$.

\subsubsection{$\Gamma= 0$, with Slow Roll}
In case all tachyons condense at once we have $\Gamma=0$ during inflation and the usual slow roll expressions apply. In order to provide a concrete example we assume $\varphi_A=\varphi_B$, tune the brane tension to $\tau_p\equiv c_1^2/(4c_2)$ and take
\begin{eqnarray}
W(\varphi)=\mathcal{N}\frac{c_1^2}{4c_2}-c_1\varphi^2+\frac{c_2}{\mathcal{N}}\varphi^4\label{potsr}
\end{eqnarray}
even  for $\varphi\sim \mathcal{O}(1)$, which is stretching the applicability of the potential \footnote{A value of $\varphi\sim 1$ corresponds to individual field values of order $\varphi_A\sim 0.1$ if $\mathcal{N}\sim \mathcal{O}(10^2)$. Since $\varphi_A\ll 1$ in order for (\ref{potsr}) to apply, we are reaching the limit of its applicability.}.

Inflation ends when either $\varepsilon$ or $\eta$ become of order one. For our potential, $\varepsilon=1$ first at
\begin{eqnarray}
\varphi_{end}=\left(4+\frac{\mathcal{N}c_1}{2c_2}-2\sqrt{4+\frac{c_1}{c_2}\mathcal{N}}\right)^{1/2}\,.
\end{eqnarray}
The value of $\varphi$ at $N=60$ efolds before the end of inflation can then be computed numerically from
\begin{eqnarray}
N\simeq -\int_{\varphi_{ini}}^{\varphi_{end}}\frac{W}{W^\prime}\,d\varphi\,.
\end{eqnarray}
Given $\varphi_{ini}$, the slow roll parameters and $n_s-1=-6\varepsilon+2\eta$ follow straightforwardly. Since the only free parameter is the number of fields, we can fine-tune $\mathcal{N}$ to yield the desired scalar spectral index. To be concrete, for $\mathcal{N}=135$ we get $\varphi_{end}\approx 15.4$ and $\varphi_{ini}\approx 4.42$ so that $\varepsilon\approx 0.0023$, $\eta\approx -0.041$ resulting in $n_s-1\approx -0.040$, matching WMAP5 \cite{Komatsu:2008hk}.

If the full tachyon potential should become steeper before $\varphi_{end}$ or the tachyons condense collectively at some smaller value $\varphi_{cond}<\varphi_{end}$, the corresponding $\varphi_{ini}$ shifts to lower values, causing $\varepsilon$ and $|\eta|$ to decrease further. As a consequence, even less fields $\mathcal{N}<135 $ are needed to match $n_{s}$.

\subsubsection{$\Gamma\neq 0$, with Slow Roll}
Based on the last two sections, we expect the contributions to $n_s$ in (\ref{finalns}) by the slow roll parameters and $\varepsilon_{\mathcal{N}}$ to be of comparable magnitude if the tachyons do not start out too close to the origin and we have $\mathcal{N}_0\sim \mathcal{O}(10^2)$ as well as a decay rate of order $\Gamma\sim \mathcal{O}(0.1)$ (or $\tilde{t}\sim \mathcal{O}(10)$). For even smaller rates, $\varepsilon_{\mathcal{N}}$ becomes negligible and the usual slow roll results apply, whereas larger values cause a premature end of slow roll inflation. On the other hand, if the tachyons start out very close to the origin, we can neglect the slow roll contribution altogether, just  as in section \ref{sec:acdwithoutslowroll}. To quantify these statements, we take the slow roll setup with $\tau_p\equiv c_1^2/(4c_2)$ as well as $\mathcal{N}_0=135$ fields and slowly turn on $\Gamma$ in order to show its effect on $n_s$.

\begin{figure}[tb]
\includegraphics[scale=0.5,angle=0]{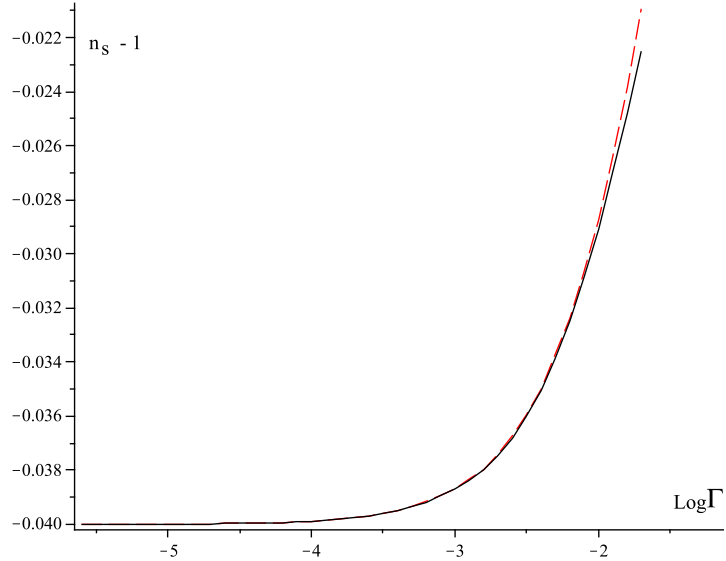}
   \caption{\label{pic:nsm1overGamma}  Setting $\tau_p\equiv c_1^2/(4c_2)$ so that the potential in (\ref{potentialacd}) vanishes at the minimum, we plot $n_s-1$ over $\log(\Gamma(t_{ini}))$, for $\Gamma=const$ (solid line) and $\Gamma=(\tilde{t}-t)^{-1}$ (dashed line). We choose $N(t_{ini})=135$, so that $n_s-1\rightarrow -0.04$ (the WMAP5 value \cite{Komatsu:2008hk}) if $\Gamma\rightarrow 0$ (or $\tilde{t}\rightarrow \infty$). As anticipated form section \ref{sec:acdwithoutslowroll}, the decay rate becomes important for $\Gamma(t_{ini})\gtrsim 10^{-2}$.}
\end{figure}

{\bf Case 1) $\Gamma=const$}\\
First, we determine the end of slow roll inflation, which occurs whenever $\varepsilon$, $\eta$ or $\varepsilon_{\mathcal{N}}$ becomes of order one (or $\mathcal{N}$ itself becomes of order one). Assuming $\mathcal{N}\gg 1$ is still valid at the end of inflation, one can show that $\varepsilon=1$ before $\eta$ and $\varepsilon_{\mathcal{N}}$ if $\Gamma<\bar{\Gamma}$, where
\begin{eqnarray}
\bar{\Gamma}\equiv -4\sqrt{\frac{3}{\mathcal{N}c_2}}\left(2-\sqrt{4+\frac{c_1}{c_2}\mathcal{N}}\right)\,.
\end{eqnarray}
Since  $\bar{\Gamma}\approx 2$ for  $\mathcal{N}=135$ fields and we are primarily interested in small decay rates, we determine $\varphi_{end}\equiv \varphi(t_{end})$ from $\varepsilon=1$. Given $\varphi_{end}$, we can determine the time $t_{ini}$, and thus the field value $\varphi(t_{ini})$ from the requirement
\begin{eqnarray}
N=\int_{t_{ini}}^{t_{end}}H\,dt\,,
\end{eqnarray}
using $H\simeq \sqrt{W/3}$ and a numerical solution to the equation of motion for $\varphi$ in (\ref{eomsrvarphi}).
 Once $\varphi_{ini}$ is known, we can straightforwardly compute $\varepsilon$, $\eta$ and $\varepsilon_{\mathcal{N}}$ at $N=60$ efolds before the end of inflation.
Using these in (\ref{finalns}), we plot the resulting spectral index over $\Gamma$  in figure \ref{pic:nsm1overGamma}. As expected, the decay rate becomes important for $\Gamma(t_{ini})\gtrsim 10^{-2}$, quickly driving the scalar spectral index outside the observationally favored region  $n_s=0.960^{+0.014}_{-0.013}$ \cite{Komatsu:2008hk} as $\Gamma$ increases further.

{\bf Case 2) $\Gamma=(\tilde{t}-t)^{-1}$}\\
We compute $n_s-1$ analogously to the previous section, but using $N(t)=N_0(1-t/\tilde{t})$ instead of $\Gamma=const$. Since $\Gamma$ is time dependent, we get a non-zero $\Gamma^\prime$, leading to
\begin{eqnarray}
\delta=\frac{2}{1-t/\tilde{t}}\frac{\sqrt{3}W^{3/2}}{\gamma W^{\prime 2}}\,,
\end{eqnarray}
which also needs to be inserted into (\ref{finalns}).
The resulting scalar spectral index is plotted over $\Gamma_{ini}\equiv \Gamma(t_{ini})=1/\tilde{t}$ in figure \ref{pic:nsm1overGamma}. $n_s$ is nearly indistinguishable from the $\Gamma=const$ case  for the small $\Gamma_{ini}$ under consideration.

\subsubsection{Summary}
Allowing for $\Gamma\neq 0$ yields the interesting possibility to drive inflation and generate observational viable values for the scalar spectral index with negligible contributions from slow roll. One may start out with a few hundred tachyons very close to the top of their potential (the potential is known in this regime). As soon as a fluctuation, be it quantum mechanical or thermal, dislodges a field, it should condense quickly and drop out of the effective description, while contributing to $\rho_r$. This is the physical origin for $\Gamma\neq 0$. If the combination $\mathcal{N}_0/\Gamma_{ini}^2$ is chosen appropriately, for instance close to $3000$ if $\Gamma=const$, then $N=60$ efolds of inflation result with $n_s-1\approx -3/N$ from (\ref{ns3overN}), in agreement with observations ($n_s-1=-0.04^{+0.014}_{-0.013}$ \cite{Komatsu:2008hk}). Admittedly, there is still some tuning involved, but it does not seem overly contrived to us: a few hundred tachyons are certainly possible and $\Gamma\sim \mathcal{O}(0.1)$ does not seem far fetched either.

If the tachyons start to roll down the potential during inflation, the usual slow roll conditions must be satisfied too, so that $n_s-1$ in (\ref{finalns}) remains small. However, the applicability of the potential in (\ref{potentialacd}) is questionable in the regime where $\varepsilon$ and $\eta$ become of interest, since $\varphi\gtrsim 1$. Further, having both, $\varepsilon$ and $\varepsilon_{\mathcal{N}}$, of similar magnitude so that their contributions to $n_s$ are comparable constitutes an additional fine-tuning, which is not needed.

We conclude that with minimal fine-tuning of $N_0\Gamma_{ini}^{-2}$ and using $\varphi_{ini}\ll 1$, the model of \cite{Majumdar:2003kd} predicts $n_s-1= - \alpha/N$, with a proportionality factor depending on the exact form of $\Gamma(t)$ (for instance $\alpha=3$ if $\Gamma=const$ and $\alpha=5/3$ if $\Gamma=(1-t/\tilde{t})^{-1}$). Thus, a constant decay rate is in good agreement with observations.

\subsection{Inflation from Multiple M5-branes \label{sec:BBK}}

In the model of \cite{Becker:2005sg} inflation is driven by an effective field $\varphi$, that evolves according to the exponential potential
\begin{eqnarray}
W(\varphi)=\tilde{W}_0e^{-\sqrt{\frac{2}{\tilde{p}}} \varphi}\,,
\end{eqnarray}
where
\begin{eqnarray}
\tilde{W}_0&=&W_0 (\mathcal{N}-1)^2\approx W_0\mathcal{N}^2\,,\\
\tilde{p}&=& \frac{\mathcal{N}(\mathcal{N}^2-1)}{c^3}\approx \left(\frac{\mathcal{N}}{c}\right)^3\,, \label{BBKp}
\end{eqnarray}
with $c=(3s\tilde{t}/4)^{1/3}\approx const \approx 19.3$  and $W_0\approx const$; here we used $\tilde{t}\approx 14$ and $s\approx 682$, as in \cite{Becker:2005sg}\footnote{Our notation differs from \cite{Becker:2005sg}, where $\tilde{t}$ is denoted by $t$.} (see also \cite{Becker:2004gw}), but other combinations are possible \cite{Ashoorioon:2006wc}.  $\mathcal{N}$ is the number of M5-branes, which are (equidistantly) stacked somewhere within an orbifold $\bf{S}^1/\bf{Z}_2$ of length $L\approx const$. Its value is constraint to $19<\mathcal{N}\ll 195$, justifying the use of $\mathcal{N}\gg 1$ above \footnote{The upper bound originates from the large volume limit invoked in \cite{Becker:2005sg}, while the lower bound guarantees $\tilde{p}>1$ (that is $\varepsilon <1 $) initially.}.

If $\mathcal{N}=const$, the scalar spectral index reads $n_s-1\simeq -2/\tilde{p}$, just like in the case of assisted inflation in section \ref{sec:staggered}. For $n_s-1=-0.04$ we need $\mathcal{N}\approx 71$ M5-branes, a number which lies well within the allowed interval. Note that this prediction is only valid as long as branes are not dissolving into the boundary branes, that is as long as the distance between adjacent branes satisfies $\Delta x\ll L/\mathcal{N}$.

In \cite{Becker:2005sg} it was also suggested that one could distribute the M5-branes uniformly over the interval in order to avoid the fine-tuning associated with the narrow stacking of branes. The number of branes is bounded by the brane separation via $\mathcal{N}\lesssim L/\Delta x$, and whenever $\mathcal{N}= L/\Delta x$ a brane dissoves into a boundary brane, decreasing $\mathcal{N}$ by one.  In terms of $\Delta x$  the effective field is defined via
\begin{eqnarray}
\varphi\equiv \tilde{t}\frac{\Delta x}{L} \sqrt{\frac{\tilde{p}}{2}} \label{BBKphi}\,,
\end{eqnarray}
where $\tilde{p}\approx (\mathcal{N}/c)^3$. Note that whenever $\mathcal{N}$ decreases, $\varphi$ makes a jump due to the decreased $\tilde{p}$, whereas $\Delta x$ is continuous (we do not perform any smoothing of $\mathcal{N}(t)$ in this section).

Let us consider the following question in order to assess whether or not we need to consider effects caused by a decreasing number of M5-branes: if we have indeed a narrow stack of $\mathcal{N}_{ini}=71$ branes initially, how many efolds follow after the outermost M5-brane hits one of the boundary branes? If the number of efolds $N_{after}$ that follow up until $\mathcal{N}\approx 19$ (at which point $\varepsilon$ becomes of order one and slow roll inflation ends) is small, the above  fixed-$\mathcal{N}$ result is applicable. On the other hand, if $N_{after}\gtrsim 60$, one cannot ignore the time dependence of $\mathcal{N}$ and cascade-inflation \cite{Ashoorioon:2006wc} results. To address this question, we assume that the stack of branes is located near the center of the orbifold, so that the branes are uniformly distributed over the whole orbifold when the first brane dissolves into a boundary brane. During slow roll, $N_{after}$ is easily computed from
\begin{eqnarray}
N_{after}\simeq -\sum_{n=19}^{\mathcal{N}_{ini}}\int_{\varphi_n^{low}}^{\varphi_{n}^{up}}\frac{W}{W^\prime}\,d\varphi\,,
\end{eqnarray}
where
\begin{eqnarray}
\frac{W}{W^{\prime}}&=&-\sqrt{\frac{\tilde{p}}{2}}=-n^{3/2}\frac{1}{\sqrt{2c^3}}
\end{eqnarray}
and the integration boundaries follow from $\varphi = \Delta x \,\tilde{t}\sqrt{\tilde{p}}/(L\sqrt{2})$ with $\Delta x^{low}=L/n$, $\Delta x^{up} = L/(n-1)$ and $\tilde{p}=n^3/c^3$ (which remains constant during each integration interval) as
\begin{eqnarray}
\varphi_n^{low}&=&\frac{1}{n}\tilde{t}\sqrt{\frac{n^3}{2c^3}} \,,\\
\varphi_n^{up}&=&\frac{1}{n-1}\tilde{t}\sqrt{\frac{n^3}{2c^3}}\,.
\end{eqnarray}
Since  the integrand is constant during each interval, we get
\begin{eqnarray}
N_{after}&\simeq&\frac{\tilde{t}}{2c^3}\sum_{n=19}^{\mathcal{N}_{ini}}n^3\left(\frac{1}{n-1}-\frac{1}{n}\right)\\
&=&\frac{\tilde{t}}{4c^3}\left[(\mathcal{N}_{ini}+1)^2+\mathcal{N}_{ini}+1+2\Psi(\mathcal{N}_{ini})-19^2-19-2\Psi(19-1)\right]\,,
\end{eqnarray}
where $\Psi$ is the digamma function (the logarithmic derivative of the gamma function). For large $\mathcal{N}_{ini}$, this can be approximated by $N_{after}\approx \tilde{t} \mathcal{N}_{ini}^2/(4c^3)=\mathcal{N}_{ini}^2/(3s)$, that is  $N_{after}\approx 2.4$ for $\mathcal{N}_{ini}=71$. Hence, inflation comes to an end shortly after the first M5-brane dissolves into the boundary brane and we are justified to use the fixed-$\mathcal{N}$ result for $n_s$. Further, since all branes dissolve within the last few efolds, we expect all of them to contribute to reheating.

One might wonder how many branes are needed to achieve $N=60$ if one starts out with a uniform distribution of M5-branes over the entire orbifold (instead of a narrow stack of branes) as mentioned in \cite{Becker:2005sg,Ashoorioon:2006wc}. Solving
\begin{eqnarray}
N&\simeq&\frac{\tilde{t}}{2c^3}\sum_{n=19}^{\mathcal{N}_{uni}}n^3\left(\frac{1}{n-1}-\frac{1}{n}\right)
\end{eqnarray}
for the number of branes yields $\mathcal{N}_{uni}\approx 350$, which is unfortunately well outside the allowed interval for $\mathcal{N}$. Thus, such an initial uniform distribution is not feasible within the framework of \cite{Becker:2005sg}, and the investigation of \cite{Ashoorioon:2006wc} (where $\mathcal{N}=66$ and  $s=682$ as well as $\tilde{t}=14$ are considered) is strictly speaking not applicable either: any imprint onto fluctuations within the last few efolds of inflation does not correspond to the large scales observed in, e.g. the CMBR. However, it should be noted that other values of $\tilde{t}$ and $s$ seem possible \cite{Ashoorioon:2006wc} so that cascade-inflation might be important in different scenarios. We leave this interesting possibility to future studies.
\footnote{Note that if  $\mathcal{N}_{uni}$ were within the allowed interval, one could not simply invoke the formalism of section \ref{sec:main}, since the potential differs from the type we considered in that it is not a simple sum, for instance $W_0\propto \mathcal{N}^2$ instead of $W_0\propto \mathcal{N}$. Therefore, one would have to adjust the formalism accordingly first.}

To summarize, in the model of \cite{Becker:2005sg} and adhering to the prefered parameter ranges, successful inflation can only be driven if the branes are stacked narrowly (this involves fine-tuning); then $n_s-1\approx 2(19.3/\mathcal{N})^3$, so that about $71$ branes are needed in order to match observations. As soon as the outermost M5-branes dissolve into the boundary branes, inflation comes to an end within a few efolds due to the rapidly decreasing number of branes. As a consequence, all branes should contribute to reheating.

\section{Conclusions \label{sec:concl}}
In this study, we investigated the consequences of disappearing inflatons during multi-field inflation. Such a process occurs quite naturally in string theoretical models of multi-field inflation, for instance if the inflatons are associated with inter-brane distances and said branes start to annihilate during inflation, or if inflatons are identified with tachyons that condense in a staggered fashion.

Focusing on inflationary models with separable potentials, we promoted the number of fields to a time dependent, continuous variable and derived the general formalism at the background and perturbed level to allow for a non-zero decay rate. This approach is only valid if the number of fields is large and the decay rate is big enough so that at least a few fields drop out of the model in a given Hubble time.  To satisfy conservation of the energy momentum tensor, an additional component to the energy budget needs to be introduced, which takes over the energy of the decaying fields. A non-zero decay rate can cause leading order corrections in observable quantities such as the scalar spectral index or the running, for which we derived general expressions. We recovered standard slow roll results and the case of a dynamically relaxing cosmological constant of  \cite{Watson:2006px} in the appropriate limits. In addition, we showed that isocurvature modes are generically suppressed during inflation for the models of interest; as a consequence, perturbations are adiabatic and nearly Gaussian. We also point out that staggered multi-field inflation might be seen as a new, less problematic type of warm inflation, which can be implemented within string theory.

Based on our formalism, we investigated three concrete models: assisted inflation with exponential potentials \cite{Liddle:1998jc}, inflation from multiple tachyons \cite{Majumdar:2003kd}, and inflation from multiple M5-branes \cite{Becker:2005sg}. For assisted inflation, the presence of a non-zero decay rate offers a graceful exit to inflation, while remaining observationally viable. For inflation from multiple tachyons, the use of our formalism offers a reliable method to compute the scalar spectral index; we concluded that the setup with best computational control, minimal fine-tuning and good agreement with observations consists of a constant decay rate slightly below one, a few hundred tachyons and a negligible slow roll contribution.  For inflation from multiple M5-branes, we showed that the dissolution of M5-branes into boundary branes is quite rapid within the allowed parameter ranges. Therefore, inflation comes to an end within a few efolds after the first M5-branes disappear. Hence, the only self consistent setup employs a (fine-tuned) narrow stack of branes, whose number is constant around the pivot scale for which the scalar spectral index is measured, and the standard slow roll result is applicable.

We did not investigate further the connection to warm inflation, since it would go beyond the scope of this article, but it is surely an avenue worth investigating more thoroughly.
Further, our selection of models is not extensive, and had the primary purpose of exemplifying the applicability of the formalism in a few concrete, distinct scenarios. Variations of the models discussed here (and also entirely different ones) are abundant in the literature and should be re-examined.

\begin{acknowledgments}
We thank Paul Steinhardt and Scott Watson for comments on the darft. D.~B. would like to thank DAMTP and A.~C.~Davis for support when this project was started, as well as Princeton University for hospitality during the final stages of this work. D.~B.~is supported by the EU FP6 Marie Curie Research and Training Network "UniverseNet" (MRTN-CT-2006-035863). T.~B.~would like to thank Kari Enquist and the Helsinki Institute of Physics for hospitality.
T.~B.~is supported by the Council on Science and Technology at Princeton University.  
This work is supported in part by PPARC grant PP/D507366/1 and the STFC at 
DAMTP.

\end{acknowledgments}

\appendix


\begin{thebibliography}{}

\bibitem{Kachru:2003sx}
  S.~Kachru, R.~Kallosh, A.~Linde, J.~M.~Maldacena, L.~P.~McAllister and S.~P.~Trivedi,
  ``Towards inflation in string theory,''
  JCAP {\bf 0310}, 013 (2003)
  [arXiv:hep-th/0308055].

\bibitem{Liddle:1998jc}
  A.~R.~Liddle, A.~Mazumdar and F.~E.~Schunck,
  ``Assisted inflation,''
  Phys.\ Rev.\  D {\bf 58}, 061301 (1998)
  [arXiv:astro-ph/9804177].

\bibitem{Malik:1998gy}
  K.~A.~Malik and D.~Wands,
  ``Dynamics of assisted inflation,''
  Phys.\ Rev.\  D {\bf 59}, 123501 (1999)
  [arXiv:astro-ph/9812204].

\bibitem{Kanti:1999vt}
  P.~Kanti and K.~A.~Olive,
  ``On the realization of assisted inflation,''
  Phys.\ Rev.\  D {\bf 60}, 043502 (1999)
  [arXiv:hep-ph/9903524].

\bibitem{Dimopoulos:2005ac}
  S.~Dimopoulos, S.~Kachru, J.~McGreevy and J.~G.~Wacker,
  ``N-flation,''
  arXiv:hep-th/0507205.

\bibitem{Calcagni:2007sb}
  G.~Calcagni and A.~R.~Liddle,
  ``Stability of multi-field cosmological solutions,''
  Phys.\ Rev.\  D {\bf 77}, 023522 (2008)
  [arXiv:0711.3360 [astro-ph]].

\bibitem{Majumdar:2003kd}
  M.~Majumdar and A.~C.~Davis,
  ``Inflation from tachyon condensation, large N effects,''
  Phys.\ Rev.\  D {\bf 69}, 103504 (2004)
  [arXiv:hep-th/0304226].

\bibitem{Becker:2005sg}
  K.~Becker, M.~Becker and A.~Krause,
  ``M-theory inflation from multi M5-brane dynamics,''
  Nucl.\ Phys.\  B {\bf 715}, 349 (2005)
  [arXiv:hep-th/0501130].


\bibitem{Ward:2007gs}
  J.~Ward,
  ``DBI N-flation,''
  JHEP {\bf 0712}, 045 (2007)
  [arXiv:0711.0760 [hep-th]].

\bibitem{Grimm:2007hs}
  T.~W.~Grimm,
  ``Axion Inflation in Type II String Theory,''
  arXiv:0710.3883 [hep-th].

\bibitem{Panigrahi:2007sq}
  K.~L.~Panigrahi and H.~Singh,
  ``Assisted Inflation from Geometric Tachyon,''
  JHEP {\bf 0711}, 017 (2007)
  [arXiv:0708.1679 [hep-th]].

\bibitem{Piao:2002vf}
  Y.~S.~Piao, R.~G.~Cai, X.~m.~Zhang and Y.~Z.~Zhang,
  ``Assisted tachyonic inflation,''
  Phys.\ Rev.\  D {\bf 66}, 121301 (2002)
  [arXiv:hep-ph/0207143].

\bibitem{Mazumdar:2001mm}
  A.~Mazumdar, S.~Panda and A.~Perez-Lorenzana,
  ``Assisted inflation via tachyon condensation,''
  Nucl.\ Phys.\  B {\bf 614}, 101 (2001)
  [arXiv:hep-ph/0107058].

\bibitem{Cline:2005ty}
  J.~M.~Cline and H.~Stoica,
  ``Multibrane inflation and dynamical flattening of the inflaton  potential,''
  Phys.\ Rev.\  D {\bf 72}, 126004 (2005)
  [arXiv:hep-th/0508029].

\bibitem{Wands:2007bd}
  D.~Wands,
  ``Multiple field inflation,''
  Lect.\ Notes Phys.\  {\bf 738}, 275 (2008)
  [arXiv:astro-ph/0702187].

\bibitem{Gordon:2000hv}
  C.~Gordon, D.~Wands, B.~A.~Bassett and R.~Maartens,
  ``Adiabatic and entropy perturbations from inflation,''
  Phys.\ Rev.\  D {\bf 63}, 023506 (2001)
  [arXiv:astro-ph/0009131].

\bibitem{Battefeld:2006sz}
  T.~Battefeld and R.~Easther,
  ``Non-gaussianities in multi-field inflation,''
  JCAP {\bf 0703}, 020 (2007)
  [arXiv:astro-ph/0610296].

\bibitem{Battefeld:2007en}
  D.~Battefeld and T.~Battefeld,
  ``Non-Gaussianities in N-flation,''
  JCAP {\bf 0705}, 012 (2007)
  [arXiv:hep-th/0703012].

\bibitem{Byrnes:2008wi}
  C.~T.~Byrnes, K.~Y.~Choi and L.~M.~H.~Hall,
  ``Conditions for large non-Gaussianity in two-field slow-roll inflation,''
  arXiv:0807.1101 [astro-ph].

\bibitem{Ashoorioon:2006wc}
  A.~Ashoorioon and A.~Krause,
  ``Power spectrum and signatures for cascade inflation,''
  arXiv:hep-th/0607001.

\bibitem{Watson:2006px}
  S.~Watson, M.~J.~Perry, G.~L.~Kane and F.~C.~Adams,
  ``Inflation without inflaton(s),''
  JCAP {\bf 0711}, 017 (2007)
  [arXiv:hep-th/0610054].

\bibitem{Chen:2006xjb}
  X.~Chen, R.~Easther and E.~A.~Lim,
  ``Large non-Gaussianities in single field inflation,''
  JCAP {\bf 0706}, 023 (2007)
  [arXiv:astro-ph/0611645].

\bibitem{Covi:2006ci}
  L.~Covi, J.~Hamann, A.~Melchiorri, A.~Slosar and I.~Sorbera,
  ``Inflation and WMAP three year data: Features have a future!,''
  Phys.\ Rev.\  D {\bf 74}, 083509 (2006)
  [arXiv:astro-ph/0606452].

\bibitem{Berera:1995ie}
  A.~Berera,
  ``Warm Inflation,''
  Phys.\ Rev.\ Lett.\  {\bf 75}, 3218 (1995)
  [arXiv:astro-ph/9509049].

\bibitem{Komatsu:2008hk}
  E.~Komatsu {\it et al.},
  ``Five-Year Wilkinson Microwave Anisotropy Probe (WMAP) Observations:
  Cosmological Interpretation,''
  arXiv:0803.0547 [astro-ph].

\bibitem{Chialva:2008zw}
  D.~Chialva and U.~H.~Danielsson,
  ``Chain inflation revisited,''
  arXiv:0804.2846 [hep-th].

\bibitem{Krause:2007jr}
  A.~Krause,
  ``Large Gravitational Waves and Lyth Bound in Multi Brane Inflation,''
  arXiv:0708.4414 [hep-th].

\bibitem{Yokoyama:1998ju}
  J.~Yokoyama and A.~D.~Linde,
  ``Is warm inflation possible?,''
  Phys.\ Rev.\  D {\bf 60}, 083509 (1999)
  [arXiv:hep-ph/9809409].

\bibitem{Mukhanov:1990me}
  V.~F.~Mukhanov, H.~A.~Feldman and R.~H.~Brandenberger,
  ``Theory of cosmological perturbations. Part 1. Classical perturbations. Part
  2. Quantum theory of perturbations. Part 3. Extensions,''
  Phys.\ Rept.\  {\bf 215}, 203 (1992).

\bibitem{Lyth:1998xn}
  D.~H.~Lyth and A.~Riotto,
  ``Particle physics models of inflation and the cosmological density
  perturbation,''
  Phys.\ Rept.\  {\bf 314}, 1 (1999)
  [arXiv:hep-ph/9807278].

\bibitem{Liddle:2000cg}
  A.~R.~Liddle and D.~H.~Lyth,
  ``Cosmological inflation and large-scale structure,''
{\it  Cambridge, UK: Univ. Pr. (2000) 400 p}

\bibitem{Kosowsky:1995aa}
  A.~Kosowsky and M.~S.~Turner,
  ``CBR anisotropy and the running of the scalar spectral index,''
  Phys.\ Rev.\  D {\bf 52}, 1739 (1995)
  [arXiv:astro-ph/9504071].

\bibitem{Malik:2002jb}
  K.~A.~Malik, D.~Wands and C.~Ungarelli,
  ``Large-scale curvature and entropy perturbations for multiple interacting
  fluids,''
  Phys.\ Rev.\  D {\bf 67}, 063516 (2003)
  [arXiv:astro-ph/0211602].

\bibitem{Kodama:1985bj}
  H.~Kodama and M.~Sasaki,
  ``Cosmological Perturbation Theory,''
  Prog.\ Theor.\ Phys.\ Suppl.\  {\bf 78}, 1 (1984).

\bibitem{Ashoorioon} A. Ashoorioon, private discussion at COSMO 2008.

\bibitem{Enqvist:2005qu}
  K.~Enqvist, A.~Jokinen, A.~Mazumdar, T.~Multamaki and A.~Vaihkonen,
  ``Non-gaussianity from instant and tachyonic preheating,''
  JCAP {\bf 0503}, 010 (2005)
  [arXiv:hep-ph/0501076].

\bibitem{Barnaby:2006cq}
  N.~Barnaby and J.~M.~Cline,
  ``Nongaussian and nonscale-invariant perturbations from tachyonic  preheating
  in hybrid inflation,''
  Phys.\ Rev.\  D {\bf 73}, 106012 (2006)
  [arXiv:astro-ph/0601481].

\bibitem{Barnaby:2006km}
  N.~Barnaby and J.~M.~Cline,
  ``Nongaussianity from tachyonic preheating in hybrid inflation,''
  Phys.\ Rev.\  D {\bf 75}, 086004 (2007)
  [arXiv:astro-ph/0611750].

\bibitem{Ohmori:2001am}
  K.~Ohmori,
  ``A review on tachyon condensation in open string field theories,''
  arXiv:hep-th/0102085.

\bibitem{Becker:2004gw}
  M.~Becker, G.~Curio and A.~Krause,
  ``De Sitter vacua from heterotic M-theory,''
  Nucl.\ Phys.\  B {\bf 693}, 223 (2004)
  [arXiv:hep-th/0403027].








\end{thebibliography}
\end{document}